\documentclass[a4paper,11pt]{article}
\usepackage{jheppub} 

\usepackage{graphicx}
\usepackage{amsmath,bm}
\usepackage{amssymb}
\usepackage{comment}
\usepackage{hyperref}

\usepackage{subfigure}
\newcommand{\bi}{\begin{itemize}}
\newcommand{\ei}{\end{itemize}}
\newcommand{\be}{\begin{enumerate}}
\newcommand{\ee}{\end{enumerate}}

\newenvironment{dfn}{{\vspace*{1ex} \noindent \bf Definition }}{\vspace*{1ex}}

\newcommand{\nn}{\nonumber}  %

\newcommand{\ket}[1]{\left| #1 \right>} 

	\newcommand{\beq}{\begin{eqnarray}}
	\newcommand{\eeq}{\end{eqnarray}}
	\newcommand{\bea}{\begin{eqnarray}\begin{aligned}}
	\newcommand{\eea}{\end{aligned}\end{eqnarray}}

\title{\boldmath Deconfined quantum criticality with internal supersymmetry}

\author[a]{Zhi-Qiang Gao,}\emailAdd{zqgao@berkeley.edu}
\author[b]{Hui Yang,}\emailAdd{huiyang.physics@gmail.com}
\author[c,1]{and Yan-Qi Wang\note{Corresponding author.}}\emailAdd{wyq@umd.edu}
\affiliation[a]{Department of Physics, University of California, Berkeley,\\
Berkeley, California 94720, USA}
\affiliation[b]{Department of Physics and Astronomy, University of Pittsburgh,\\
Pittsburgh, Pennsylvania 15213, USA}
\affiliation[c]{Department of Physics and Joint Quantum Institute, University of Maryland,\\
College Park, Maryland 20742, USA}
\abstract{
Deconfined quantum critical point (DQCP) describes direct, non-fine-tuned quantum phase transition between two ordered phases that break distinct and seemingly unrelated symmetries, providing a route to continuous phase transition beyond the conventional Ginzburg--Landau paradigm. In this work we extend the DQCP paradigm to systems with \emph{internal} supersymmetry (SUSY), where the on-site Hilbert space furnishes a representation of a Lie superalgebra, and the Hamiltonian is invariant under the corresponding Lie supergroup. Focusing on the minimal supersymmetric generalization of spin $SU(2)$, namely $OSp(1|2)$, we propose a supersymmetric deconfined quantum critical point (sDQCP) between a phase that breaks internal $OSp(1|2)$ and a phase that instead breaks lattice rotation symmetry. We formulate a non-linear sigma model on the supersphere target space that captures the symmetry intertwinement characteristic of the sDQCP, and we further develop a gauge theory description to address its dynamical properties, including an argument for 3D XY critical behavior. Finally, we show that explicitly breaking $OSp(1|2)$ down to $SU(2)$ continuously connects our sDQCP to the conventional DQCP scenario.
}
\begin{document}
\maketitle
\flushbottom

\section{Introduction}
\label{sec:intro}

Deconfined quantum critical point (DQCP)~\cite{Senthil:2004aa,Senthil:2004ab,Grover2007,Grover2008,Wang2015,Roberts2019,Zhang2023,Cui2025} marks unconventional quantum critical point~\cite{Xu:2012} beyond the traditional Landau paradigm~\cite{Landau}. It describes direct, non-fine-tuned quantum phase transition between two distinct ordered phases characterized by different symmetry breaking patterns. More precisely, let $G$ denote the parent symmetry group, which is spontaneously broken to $H_1$ in phase~I and to $H_2$ in phase~II. By ``different'' symmetry breaking patterns we mean that neither $H_1$ is a subgroup of $H_2$ nor $H_2$ a subgroup of $H_1$, so that phases~I and~II correspond to distinct ordered phases. If a direct quantum phase transition between these two phases exists without fine tuning, it is described by a DQCP. 

\begin{figure}[htbp]
\centering
\includegraphics[width=0.3\linewidth]{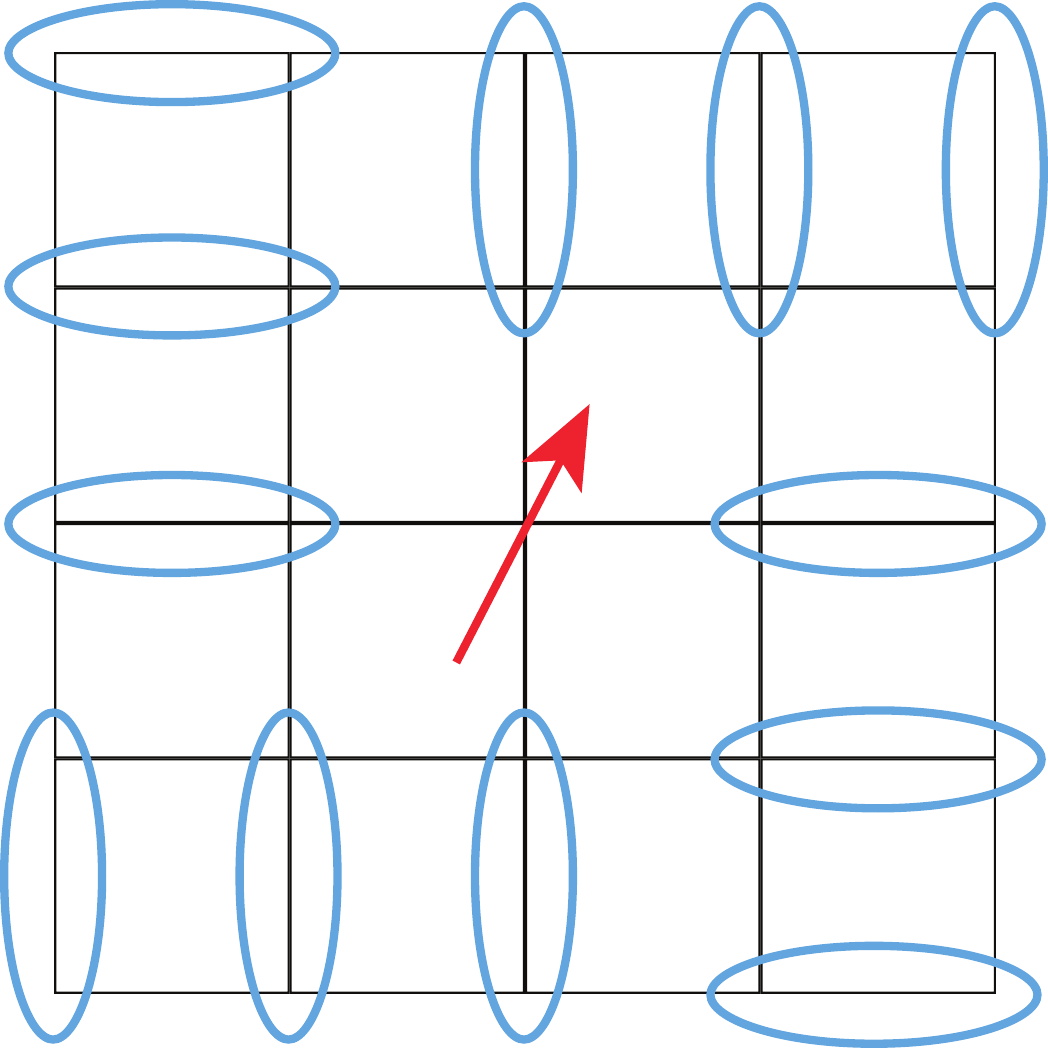}
\caption{Illustration of a VBS (SVBS) vortex. Spin singlets formed by spins on nearest-neighbor bonds are denoted as blue ellipses. The vortex core, which carries spin-$\frac{1}{2}$, is denoted as the red arrow. Around this VBS (SVBS) vortex the four-fold lattice rotation symmetry is locally restored.
}
\label{fig:VBS}
\end{figure}

The kinetics of DQCP is encoded in the non-trivial interplay between the two unbroken symmetries, $H_1$ and $H_2$, that a defect or texture of one symmetry carries the charge of the other. This signals a mixed anomaly between $H_1$ and $H_2$~\cite{Wang:2017uq,Metlitski2018,Lu2023,Wang2024,Pace2024}. Consequently, the proliferation of such defects or textures of one symmetry, while restoring this symmetry, spontaneously breaks the other. To be concrete, consider the N\'eel to valence bond solid (VBS) transition on two dimensional square lattice, the prototype of DQCP~\cite{Senthil:2004aa,Senthil:2004ab}. At the Hamiltonian level, the system has both spin rotation symmetry $SU(2)_S$ and lattice rotation symmetry $(\mathbb{Z}_4)_R$, while the latter will be promoted to $U(1)_R$ close to the phase transition point since the four-fold rotational symmetry breaking is irrelevant in (2+1)D. Here subscript $S$ and $R$ denotes spin and lattice rotation, respectively. The N\'eel phase breaks $SU(2)_S$ but preserves $U(1)_R$, while the VBS phase breaks $U(1)_R$ but preserves $SU(2)_S$. In the VBS phase, each pair of spins sitting on two nearest-neighbored sites form a $SU(2)_S$ singlet. A defect in this phase is a VBS vortex, around which the $(\mathbb{Z}_4)_R$ lattice rotation symmetry is locally restored, as shown in figure~\ref{fig:VBS}. However, since the site at the VBS vortex core is not bonded with any other sites to form a spin singlet, it carries a spin-$\frac{1}{2}$ under $SU(2)_S$, {\it i.e.,} the VBS vortex is charged under $SU(2)_S$. Therefore, when VBS vortices proliferate and restore $(\mathbb{Z}_4)_R$, it will spontaneously break $SU(2)_S$ and drive the system into the N\'eel phase. Similarly, in the N\'eel phase the textures are skyrmions which carry lattice angular momenta, {\it i.e.,} charge of $(\mathbb{Z}_4)_R$. Therefore, proliferation of skyrmions restore $SU(2)_S$ but spontaneously breaks $(\mathbb{Z}_4)_R$, driving the system into the VBS phase.

The dynamics of DQCP is more complicated. Whether the N\'eel to VBS transition is indeed a continuous transition or a weakly first order transition is still under debate~\cite{Sandvik:2007aa,Kuklov:2008aa,Wang:2017uq,li2019,Huang2019,Zhao2022,Liu2023,Song2025}. A theoretical argument of the transition to be continuous is based on the non-compact $\mathbb{CP}^1$ model~\cite{Senthil:2004aa,Motrunich:2004}. Here non-compact means monopoles are suppressed in the $U(1)$ gauge field. The suppression of monopoles arises from the lattice geometry that restricts the skyrmion number in the N\'eel order can only change by multiples of four~\cite{Haldane:1988}, suggesting the monopole events be quadrupoled and hence irrelevant at the (2+1)D critical point~\cite{Senthil:2004aa,Senthil:2004ab}. Therefore, in the vicinity of the critical point, the spinons coupled to the non-compact $U(1)$ gauge field are asymptotically deconfined.

Supersymmetry (SUSY) is a proposed extension of spacetime symmetry that relates bosons and fermions within a unified framework, originally proposed as a generalization of Poincar\'e symmetry~\cite{Golfand1971,Wess1974,Freedman1976,Deser1976} and a solution to the gauge hierarchy problem~\cite{Wess1974,Fayet1976,Fayet1977,Sakai1981}. Conceptually, SUSY posits that the fundamental degrees of freedom come in paired {\it superpartner} multiplets, so that transformations can exchange fermionic and bosonic states while preserving the underlying dynamics. If realized in nature either exactly or as an approximate symmetry emergent in certain regimes, SUSY has far-reaching consequences, that it can lead to improved theoretical control over quantum corrections, enable deeper connections between seemingly different theories, and provide a powerful organizing principle for constructing and constraining models of physics beyond the Standard Model. Generalization of spacetime SUSY includes quantum mechanical SUSY and internal SUSY. In quantum mechanical SUSY, the Hamiltonian is given by the anti-commutator of two fermionic operators. Since Hamiltonian is the time-component of the momentum 4-vector, quantum mechanical SUSY can be viewed as {\it time-direction SUSY} which is still related to spacetime. In the spectrum of a Hamiltonian with quantum mechanical SUSY, each bosonic (fermionic) excited state with even (odd) fermion parity has a fermionic (bosonic) partner~\cite{Fendley:2003aa,Yang:2004}, rendering the SUSY nature of this Hamiltonian. Typically, the low energy effective theories of quantum mechanical SUSY models have emergent spacetime SUSY, even if the quantum mechanical SUSY is not exact in UV~\cite{Fendley:2003aa,Yang:2004,OBrien:2018aa,Gao:2020,Gao2024}. Another generalization of SUSY acts as an internal symmetry, which is not related to space and time. On a lattice, internal SUSY means that on each site the local Hilbert space spans a representation of some Lie superalgebra~\cite{Frappat:1996}, a generalization of Lie algebra that contains fermionic generators, as conserved quantities of the system. Consequently, the Hamiltonian of such lattice system will have the Lie supergroup symmetry corresponding to the Lie superalgebra. 

The fermionic generators of the Lie superalgebra have anticommutation relations. In this work, we mainly focus on the $OSp(1|2)$ Lie supergroup symmetry, which is an $\mathcal{N}=1$ internal SUSY generalization of $SU(2)$. Its five generators of $OSp(1|2)$, $S_{a=1,2,3}$ and $V_{\alpha=1,2}$ satisfy commutation and anticommutation relations~\cite{Frappat:1996}
\beq
\left[S_a,S_b\right]=i\epsilon_{abc}S_c,\quad\left[S_a,V_\alpha\right]=\frac{1}{2}\left(\sigma_a\right)_{\beta\alpha}V_\beta,\quad\left\{V_\alpha,V_\beta\right\}=\frac{1}{2}\left(J\sigma_a\right)_{\alpha\beta}S_a,\label{eq:5}
\eeq
where $\sigma_{a=1,2,3}$ are Pauli matrices and $J=i\sigma_2$. $S_a$ generate the $SU(2)$ subgroup of $OSp(1|2)$, and $V_\alpha$ form a spin-$1/2$ irrep of this $SU(2)$. Bosonic generators $S_a$ remain Hermitian, while fermionic generators $V_\alpha$ satisfying anticommutation relations are non-Hermitian. Similar to $SU(2)$, irreps of $OSp(1|2)$ can be also labeled by an integer or half-integer $S$, which is called {\it spin} in parallel of $SU(2)$. The dimension of a spin-$S$ irrep of $OSp(1|2)$ is $(4S+1)$. The generators of $OSp(1|2)$ under its smallest non-trivial irrep (three dimensional with spin $S=\frac{1}{2}$) read
\beq
S_a=\frac{1}{2}\begin{pmatrix}
\sigma_a & 0\\
0 & 0
\end{pmatrix},\quad
V_\alpha=\frac{1}{2}\begin{pmatrix}
0 & \tau_\alpha\\
-\left(J\tau_\alpha\right)^\mathbf{T} & 0
\end{pmatrix},\label{eq:6}
\eeq
where $\tau_1=(1,0)^\mathbf{T}$ and $\tau_2=(0,1)^\mathbf{T}$ are eigenvectors of $\sigma_3$ with eigenvalue $\pm 1$ ({\it i.e.} $SU(2)$ spin up and down). The non-Hermiticity of $V_\alpha$ is clear in (\ref{eq:6}). The Casimir operator of $OSp(1|2)$ is defined as $C=S_aS_a+V_\alpha J_{\alpha\beta}V_\beta$, which is equal to $S(S+\frac{1}{2})$ for spin-$S$ irrep.

In this work, we extend the DQCP paradigm to systems with internal SUSY. In Sec.~\ref{sec:lattice}, we introduce a lattice model whose on-site Hilbert space transforms as spin-$\frac{1}{2}$ under the Lie supergroup $OSp(1|2)$~\cite{Frappat:1996}, and we present an $OSp(1|2)$-symmetric Hamiltonian. In Sec.~\ref{sec:SVBS}, we discuss the super-VBS (SVBS) phase, which breaks lattice rotation symmetry while preserving $OSp(1|2)$. In Sec.~\ref{sec:SN}, we introduce the super-N\'eel (SN) phase, which breaks $OSp(1|2)$ while preserving lattice rotation symmetry. In Sec.~\ref{sec:kin}, we formulate a non-linear sigma model with a supersphere target space that captures the kinetics of the sDQCP, \emph{i.e.}, the intertwinement between lattice symmetry and internal SUSY. In Sec.~\ref{sec:dyn}, we further develop a gauge theory description to address the dynamical properties of the transition, including an argument for 3D XY critical behavior. In Sec.~\ref{sec:break}, we show that explicitly breaking $OSp(1|2)$ down to $SU(2)$ continuously connects our sDQCP to the conventional DQCP scenario~\cite{Senthil:2004aa,Senthil:2004ab}. Finally, in Sec.~\ref{sec:con}, we conclude and outline future directions.

\section{The model}\label{sec:model}

In this section, we introduce the system potentially hosting the sDQCP. In Sec.~\ref{sec:lattice} we define the lattice Hamiltonian that exhibits super-VBS (SVBS) and super-N\'eel (SN) ground states depending on tuning parameters. In Sec.~\ref{sec:SVBS} and \ref{sec:SN} we discuss in detail the definitions, symmetry breaking patterns and symmetry defects of SVBS and SN states, respectively.

\subsection{The lattice Hamiltonian}\label{sec:lattice}

To accommodate internal SUSY, we consider a two dimensional square lattice, where each site spans a three dimensional Hilbert space hosting an $OSp(1|2)$ spin-$\frac{1}{2}$. The onsite commutation relations of $OSp(1|2)$ generators read,
\begin{subequations}\label{eq:7}
\begin{align}
&\left[S_a(\mathbf{i}),S_b(\mathbf{j})\right]=i\delta_{\mathbf{i}\mathbf{j}}\epsilon_{abc}S_c(\mathbf{i}),\\
&\left[S_a(\mathbf{i}),V_\alpha(\mathbf{j})\right]=\frac{1}{2}\delta_{\mathbf{i}\mathbf{j}}\left(\sigma_a\right)_{\beta\alpha}V_\beta(\mathbf{i}),\label{eq:7}\\
&\left\{V_\alpha(\mathbf{i}),V_\beta(\mathbf{j})\right\}=\frac{1}{2}\delta_{\mathbf{i}\mathbf{j}}\left(J\sigma_a\right)_{\alpha\beta}S_a(\mathbf{i}).
\end{align}
\end{subequations}
Note that $V_\alpha(\mathbf{i})$ on different lattice sites anti-commute with each other, suggesting its fermionic nature. We define the two-site Casimir operator $C({\bf ij})$ as~\cite{Arovas:2009}
\beq
C(\mathbf{i}\mathbf{j})=S_a(\mathbf{i})S_a(\mathbf{j})+V_\alpha(\mathbf{i})J_{\alpha\beta}V_\beta(\mathbf{j}),\label{eq:8}
\eeq
which is invariant under $OSp(1|2)$ operations. The lattice Hamiltonian $H$ with an internal $OSp(1|2)$ symmetry is defined through a polynomial of two-site Casimir operators $C({\bf ij})$:
\beq
H=K\sum_{\left<\mathbf{i}\mathbf{j}\right>}C(\mathbf{i}\mathbf{j})+H^\prime[C({\bf ij})],\label{eq:9}
\eeq
where the summation of $C(\mathbf{i}\mathbf{j})$ on nearest-neighbor sites $\left<\mathbf{i}\mathbf{j}\right>$ resembles an anti-ferromagnetic (AFM) Heisenberg-type interaction with positive coupling constant $K$, and $H^\prime$ includes higher order interactions of $C(\mathbf{i}\mathbf{j})$. The Hamiltonian (\ref{eq:9}) exhibits different phases, including both SN and SVBS as ground states, via adjusting the form of $H^\prime$~\cite{Arovas:2009,Hasebe:2013}. We will discuss the details of the SVBS phase in Sec.~\ref{sec:SVBS}, and the SN phase in Sec.~\ref{sec:SN}.

It is crucial to notice that the two-site Casimir operator (\ref{eq:8}) is non-Hermitian. More generally, lattice Hamiltonians with internal SUSY are typically non-Hermitian but pseudo-Hermitian~\cite{Saleur:2003,Arovas:2009,Hasebe:2013,Frahm:2022}. A pseudo-Hermitian Hamiltonian $H$ satisfies $H^\dagger=PHP$ for some unitary and Hermitian operator $P$. Pseudo-Hermitian Hamiltonian was first introduced in Ref.~\cite{Pauli:1943} and closely related to $\mathcal{P}\mathcal{T}$-symmetric Hamiltonian widely studied in non-Hermitian systems~\cite{Bender:1999}. Such a Hamiltonian has a real spectrum~\cite{Pauli:1943}, with a well defined unitary time evolution~\cite{Pauli:1943,Mostafazadeh:2002}.  For (\ref{eq:9}), the operator $P$ reads,
\beq
P=\prod_\mathbf{i}P(\mathbf{i}),\quad P(\mathbf{i})=\begin{pmatrix}
\sigma_2 & 0\\
0 & 1
\end{pmatrix},\quad [P(\mathbf{i}),P(\mathbf{j})]=0.\label{eq:10}
\eeq
Each onsite $P(\mathbf{i})$ acts on the onsite $OSp(1|2)$ generator as $P(\mathbf{i})S_a(\mathbf{i})P(\mathbf{i})=-S_a^\mathbf{T}(\mathbf{i})$ and $P(\mathbf{i})V_\alpha(\mathbf{i})P(\mathbf{i})=-iV^\dagger_\alpha(\mathbf{i})$. Therefore $C({\bf ij})$ satisfies the  $P(\mathbf{i})P(\mathbf{j})C(\mathbf{i}\mathbf{j})P(\mathbf{i})P(\mathbf{j})=C^\dagger(\mathbf{i}\mathbf{j})$ pseudo-Hermiticity, and so as the Hamiltonian (\ref{eq:9}).  

\subsection{The super-VBS phase}\label{sec:SVBS}

In the SVBS phase of Hamiltonian (\ref{eq:9}), the lattice rotation symmetry $(\mathbb{Z}_4)_R$ is spontaneously broken, while the internal $OSp(1|2)$ symmetry is preserved. Similar to the symmetry breaking pattern in the usual VBS phase~\cite{Senthil:2004aa,Senthil:2004ab}, the Goldstone manifold of the SVBS phase is also parameterized by VBS order parameters $v_1$ and $v_2$ with $v_1^2+v_2^2=1$. An SVBS vortex sits on a dangling site around which the lattice rotation symmetry is locally restored. Since each site carries a spin-$\frac{1}{2}$ irrep of $OSp(1|2)$, such an SVBS vortex carries $OSp(1|2)$ spin-$\frac{1}{2}$ as well.

The ground state wavefunction of the SVBS phase can be formulated by a parton construction~\cite{Arovas:2009}. In this parton theory two bosonic partons created by $b^\dagger_{1,2}(\mathbf{i})$ and one fermionic parton created by $f^\dagger(\mathbf{i})$ are introduced on each lattice site $\mathbf{i}$, with the onsite-Hilbert space constraint
\beq
b^\dagger_{1}(\mathbf{i})b_{1}(\mathbf{i})+b^\dagger_{2}(\mathbf{i})b_{2}(\mathbf{i})+f^\dagger(\mathbf{i})f(\mathbf{i})=1,\label{eq:101}
\eeq
which introduces a $U(1)$ gauge constraint. The $OSp(1|2)$ generators are constructed from the three-component spinor $\psi^\dagger(\mathbf{i})=(b^\dagger_1(\mathbf{i}), b^\dagger_2(\mathbf{i}), f^\dagger(\mathbf{i}))$ as~\cite{Arovas:2009} $S_a(\mathbf{i})=\psi^\dagger(\mathbf{i})S_a\psi(\mathbf{i})$ and $ V_\alpha(\mathbf{i})=\psi^\dagger(\mathbf{i})V_\alpha\psi(\mathbf{i})$, where $S_a$ and $V_\alpha$ are defined in (\ref{eq:6}). As illustrated in figure~\ref{fig:VBS}, the SVBS ground state is created by the production of operators $\chi^\dagger(\mathbf{i}\mathbf{j})$ on lattice bonds $\left<\mathbf{i}\mathbf{j}\right>$ circled by blue ellipse
\beq
\ket{\mathrm{SVBS}}=\prod_{\text{circled }\left<\mathbf{i}\mathbf{j}\right>}\chi^\dagger(\mathbf{i}\mathbf{j})\ket{\mathrm{vac}},\label{eq:103}
\eeq
with $OSp(1|2)$ singlet $\chi^\dagger(\mathbf{i}\mathbf{j})=b^\dagger_1(\mathbf{i})b^\dagger_2(\mathbf{j})-b^\dagger_2(\mathbf{i})b^\dagger_1(\mathbf{j})+f^\dagger(\mathbf{i})f^\dagger(\mathbf{j})$~\citep{Arovas:2009}.

\subsection{The super-N\'eel phase}\label{sec:SN}

In the SN phase of Hamiltonian (\ref{eq:9}), $\mathbf{S}(\mathbf{i})$ is condensed, while $V_\alpha(\mathbf{i})$ cannot be condensed due to its fermionic nature. The $OSp(1|2)$ symmetry is spontaneously broken to $U(1)$, leaving a supersphere Goldstone manifold $OSp(1|2)/U(1)= S^{2|2}$. A unit supersphere $S^{p|2}$ is a supermanifold~\cite{Rabin:1985} parameterized by $(p+1)$ bosonic coordinates $x_{i=1,2,\cdots,p+1}$ and 2 fermionic coordinates $\theta_{\nu=1,2}$ satisfying $x_ix_i+\theta_\nu J_{\nu\rho}\theta_\rho =1$.
Here $\theta_\nu$ are Grassmann numbers with $\theta_1\theta_2=-\theta_2\theta_1$ and $\theta_1\theta_1=\theta_2\theta_2=0$. We further define $\hat{x}_i=x_i(1+\theta_1\theta_2)$ which parameterizes a unit sphere $S^p$. This unit sphere $S^p$ is called the body of $S^{p|2}$, and the rest fermionic coordinates $\theta_1$ and $\theta_2$ are called the soul~\cite{Rabin:1985}. Supersphere are homotopically equivalent to its body, {\it i.e.}, $\pi_q(S^{p|2})=\pi_q(S^p)$~\cite{Rabin:1985}.

The NL$\sigma$M describing the Goldstone modes of the SN phase reads
\beq
S=\frac{1}{2g^2}\int_{S^{2|2}}\mathrm{d}^3x~\left(\partial_\mu n_a\partial_\mu n_a+\partial_\mu\eta_\alpha J_{\alpha\beta}\partial_\mu \eta_\beta\right),\label{eq:14}
\eeq
with $n_an_a+\eta_\alpha J_{\alpha\beta}\eta_\beta=1$. The gapless bosonic modes $n_a$ are identified with the condensate $S_a(\mathbf{i})$ via $n_a=\left<(-1)^\mathbf{i}S_a(\mathbf{i})\right>$. The two gapless fermionic modes $\eta_\alpha$ are corresponding to $V_\alpha(\mathbf{i})$ and play the role of {\it Goldstino}, the SUSY partners of Goldstone bosons. The superskyrmion number in the SN phase characterized by $\pi_2(S^{2|2})=\mathbb{Z}$ is
\beq
Q=\frac{1}{8\pi}\int_{S^2}\epsilon_{abc}\hat{n}_a\mathrm{d}\hat{n}_b\mathrm{d}\hat{n}_c=\frac{1}{8\pi}\int_{S^{2|2}}\epsilon_{abc}n_a\mathrm{d}n_b\mathrm{d}n_c\left(1+\frac{3}{2}\eta_\alpha J_{\alpha\beta}\eta_{\beta}\right),\label{eq:15}
\eeq
where $\hat{n}_a=n_a(1+\eta_1\eta_2)$ parameterizes the body of $S^{2|2}$. The statement, that the superskyrmion number can change only in multiples of four, continues to hold in the SN phase, as in the conventional N\'eel phase. It depends only on the lattice geometry and on the quantization of the soliton number~\cite{Haldane:1988}. Therefore, supermonopoles~\cite{Hasebe:2005} must be also quadrupoled (grouped in four), similar to the monopoles in the N\'eel phase~\cite{Haldane:1988}.

\section{Kinetics: Non-linear sigma model formalism}\label{sec:kin}

In analogue to the $O(5)$ NL$\sigma$M description~\cite{Grover2008} of DQCP, we develop an NL$\sigma$M with a level-1 WZW term to describe the kinetics of the sDQCP between SN and SVBS phases. The Goldstone modes ${\bm w}=(n_1,n_2,n_3,v_1,v_2)$ are unified with the two Goldstino modes $\eta_{1,2}$ as $w_aw_a+\eta_\alpha J_{\alpha\beta}\eta_{\beta}=1$, parametrizing the unit supersphere~\cite{Rabin:1985} $S^{4|2}$ target manifold. The NL$\sigma$M reads
\beq
S&=&\frac{1}{2g^2}\int_{S^{4|2}}\mathrm{d}^3x\,(\partial_\mu w_a\partial_\mu w_a+\partial_\mu\eta_\alpha J_{\alpha\beta}\partial_\mu \eta_\beta)\nn\\
&&-\frac{2\pi i}{64\pi^2}\int_{\mathcal{M}}\epsilon_{abcde}\tilde{w}_a\mathrm{d}\tilde{w}_b\mathrm{d}\tilde{w}_c\mathrm{d}\tilde{w}_d\mathrm{d}\tilde{w}_e\left(1+\frac{5}{2}\tilde{\eta}_\alpha J_{\alpha\beta}\tilde{\eta}_{\beta}\right).\label{eq:20}
\eeq
Here the target manifold of the WZW term, $\mathcal{M}$, is the extension of $S^{4|2}$ with $\partial\mathcal{M}=S^{4|2}$~\cite{CFT}. Field variables $\tilde{w}_a$ and $\tilde{\eta}_a$ represent a one-parameter family extension of the field configuration $w_a$ and $\eta_\alpha$ to a trivial configuration, such that $\tilde{w}_a(x,y,t,u=0)=w_a(x,y,t)$, $\tilde{\eta}_\alpha(x,y,t,u=0)=\eta_\alpha(x,y,t)$, and $\tilde{w}_a(x,y,t,u=1)=\delta_{a5}$, $\tilde{\eta}_\alpha(x,y,t,u=1)=0$. This extension exists since $\pi_3(S^{4|2})=\{0\}$. 

In what follows we show that an SVBS vortex indeed carries spin-$\frac{1}{2}$ under $OSp(1|2)$ from the WZW term defined in (\ref{eq:20})~\cite{Huang:2023}. Consider an SVBS vortex loop in (2+1)D spacetime. Away from the vortex core, the system is deep in the SVBS phase, where $v_1^2+v_2^2\rightarrow 1$ and $n_1^2+n_2^2+n_3^2+\eta_1\eta_2-\eta_2\eta_1\rightarrow 0$. Close to the vortex core, the SVBS order is locally destroyed, suggesting $v_1^2+v_2^2\rightarrow 0$ and $n_1^2+n_2^2+n_3^2+\eta_1\eta_2-\eta_2\eta_1\rightarrow 1$. Consequently, the field configuration of a vortex loop can be parameterized as
\beq
{\bm w}(r,\varphi,t)=\left(\sqrt{1-h(r)^2}{\bm n}(t),h(r)\cos \varphi,h(r)\sin \varphi\right),\label{eq:23}
\eeq
where $r$ and $\varphi$ are polar coordinates measured from the vortex loop. Function $h(r)$ is chosen to have $h(r)\rightarrow 1$ for $r\rightarrow 0$ and $h(r)\rightarrow 0$ for $r\rightarrow +\infty$. To satisfy $n_1^2+n_2^2+n_3^2+\eta_1\eta_2-\eta_2\eta_1=1$, fermionic fields $\eta_\alpha$ should be redefined as $\sqrt{1-h(r)^2}\eta_\alpha$. 
Since $\pi_1(S^{2|2})=\{0\}$, we can extend the field configuration in $u$ coordinate by deforming $n_a$ and $\eta_\alpha$ to have $\tilde{n}_a(t,u=0)=n_a(t)$, $\tilde{\eta}_\alpha(t,u=0)=\eta_\alpha(t)$ and $\tilde{n}_a(t,u=1)=\delta_{a3}$, $\tilde{\eta}_\alpha(t,u=1)=0$. Plugging
\begin{subequations}
    \begin{align}
        &\tilde{{\bm w}}(r,\varphi,t,u)=\left(\sqrt{1-h(r)^2}\tilde{{\bm n}}(t,u),h(r)\cos \varphi,h(r)\sin \varphi\right),\\
        &\tilde{\eta}_\alpha(t,u)\mapsto \sqrt{1-h(r)^2}\tilde{\eta}_\alpha(t,u),
    \end{align}
\end{subequations}
into (\ref{eq:20}) and integrating over $r,\varphi$ reduces the WZW term to
\beq
S_\mathrm{WZW}=-\frac{2\pi i}{4\pi}\int_{\mathcal{D}}\mathrm{d}t\mathrm{d}u~\epsilon_{abc}\tilde{n}_a\partial_t\tilde{n}_b\partial_u\tilde{n}_c\left(1+\frac{3}{2}\tilde{\eta}_\alpha J_{\alpha\beta}\tilde{\eta}_{\beta}\right),\label{eq:25}
\eeq
where the integration is conducted on target manifold $\mathcal{D}$ with $\partial\mathcal{D}=S^{2|2}$. This is exactly the Berry phase of an $OSp(1|2)$ spin-$\frac{1}{2}$ in (0+1)D. Therefore, we conclude that the SVBS vortex carries $OSp(1|2)$ spin-$\frac{1}{2}$, in accordance with the physical picture of SVBS vortices.

\section{Dynamics: Gauge theory formalism}\label{sec:dyn}

The NL$\sigma$M formalism captures the kinetics of the sDQCP about intertwinement of symmetry defects and symmetry charges. To investigate the dynamical aspects such as critical phenomena, we turn to a gauge theory which is a SUSY generalization of the original proposal~\cite{Senthil:2004aa,Senthil:2004ab,Motrunich:2004}.

The unit supersphere $S^{2|2}$ can be parameterized by two complex bosonic coordinates $z_1,z_2$ and one complex fermionic coordinate $\xi$ as~\cite{Read:2001,Hasebe:2005,Kurkcuoglu:2004}
\beq
n_a=\bar{\Psi}S_a\Psi,\quad\eta_\alpha=\bar{\Psi}V_\alpha\Psi.\label{eq:26}
\eeq
Here the field $\Psi=(z_1,z_2,\xi)^\mathbf{T}$ with $\bar{\Psi}=(\bar{z}_1,\bar{z}_2,-\bar{\xi})$ is a spin-$\frac{1}{2}$ spinor of $OSp(1|2)$, and $\bar{z}_{1,2}$ is the ordinary complex conjugate of $z_{1,2}$. For a complex Grassmann number $\xi=\vartheta_1+i\vartheta_2$ where real Grassmann number $\vartheta_{1,2}$ represent its real and imaginary part respectively, $\bar{\xi}$ is defined as $\bar{\xi}=\vartheta_2+i\vartheta_1$. Therefore, the normalization of $n_an_a+\eta_\alpha J_{\alpha\beta}\eta_\beta=1$ manifests $\bar{\Psi}\Psi=1$. The definition of $\Psi$ has a $U(1)$ phase redundancy, such that $\Psi\mapsto e^{i\phi}\Psi$, $\phi\in [0,2\pi)$, which leaves (\ref{eq:26}) unchanged. Upon gauging this $U(1)$ redundancy, $\Psi$ parameterizes $OSp(1|2)/U(1)=S^{2|2}$, or equivalently $S^{3|2}/S^1=\mathbb{CP}^{1|1}$, which is the Goldstone manifold of the SN phase. This is also consistent with the parton construction~\cite{Arovas:2009} of the SVBS phase in Sec.~\ref{sec:SVBS}, where $z_{1,2}$ and $\xi$ are identified as $b_{1,2}$ and $f$, respectively. The $U(1)$ gauge constraint arising in (\ref{eq:101}) as a local charge conservation is naturally identified as the $U(1)$ phase redundancy in $\Psi$.

This gauging procedure can be seen by plugging (\ref{eq:26}) into (\ref{eq:14}). The resulting action becomes
\beq
S=\frac{1}{2g^2}\int_{S^{2|2}}\mathrm{d}^3x~(\partial_\mu+ia_\mu)\bar{\Psi}(\partial_\mu-ia_\mu)\Psi,\label{eq:27}
\eeq
where $a_\mu$ is a dynamical $U(1)$ gauge field whose equation of motion yields~\cite{Kurkcuoglu:2004} $a_\mu=\frac{i}{2}(\bar{\Psi}\partial_\mu\Psi-(\partial_\mu\bar{\Psi})\Psi)=\partial_\mu \phi$. The flux quanta $\Phi$ of $a_\mu$ is related to the superskyrmion number defined in (\ref{eq:15}) by $\Phi=\frac{1}{2\pi}\int_{S^{2|2}}\mathrm{d}a=Q$~\cite{Landi:2001,Hasebe:2005},
which is conserved upon modulo 4~\cite{Haldane:1988}. Therefore, with the suppression of supermonopoles, this $U(1)$ gauge field becomes non-compact. By softening the normalization of $\Psi$ and including a Maxwell term of $a_\mu$ in the vicinity of the critical point, we obtain the following Lagrangian
\beq
\mathcal{L}&=&\sum_{\alpha=1,2} \left|\left(\partial_\mu-ia_\mu\right)z_\alpha\right|^2+s\left|z\right|^2+u\left|z\right|^4+\frac{1}{2\kappa}\left(\epsilon_{\mu\nu\rho}\partial_\nu a_\rho\right)^2\nn\\
&&+(\partial_\mu-ia_\mu)\xi(\partial_\mu+ia_\mu)\bar{\xi}+s\xi\bar{\xi}+2u\left|z\right|^2\xi\bar{\xi},\label{eq:30}
\eeq
where $s$ denotes the mass of complex boson field $z$ and symplectic fermion field $\xi$, $u>0$ represents the self-interaction of $z$, and $\kappa>0$ is the Maxwell coupling of $a_\mu$. The first line of (\ref{eq:30}) is the standard $\mathbb{CP}^1$ model, where the two-component complex boson field $z$ is coupled to a dynamical $U(1)$ gauge field. The second line of (\ref{eq:30}) describes the interactions between the symplectic fermion field $\xi$ and the $U(1)$ gauge field $a_\mu$ as well as the boson field $z$. The symplectic fermion has second order derivatives of spacetime in its equation of motion, same as the complex boson. In fact, this is enforced by the internal SUSY that rotates the spinor $\Psi$ via $S_a$ and $V_\alpha$ in (\ref{eq:6}). An interacting symplectic fermion field theory is also pseudo-unitary~\cite{LeClair:2006,LeClair:2007,Kapit:2009,Robinson:2009}, in accordance with the pseudo-Hermiticity of the $OSp(1|2)$ symmetric lattice Hamiltonian. 

Phases and phase transitions can be analyzed via (\ref{eq:30}). For $s>0$, both $z$ and $\xi$ are gapped, and the internal $OSp(1|2)$ symmetry is preserved. Their masses are equal to each other $m_z^2=m^2_\xi=s$ as required by the internal SUSY. The $U(1)$ gauge field is in its Coulomb phase, with a gapless dual photon excitation. Approaching the critical point, the mass of $z$ and $\xi$ decreases, and the spinons are asymptotically deconfined. The quadrupoled supermonopoles will drive the critical spin liquid into the SVBS phase, where both the $U(1)$ gauge field and the spinon fields become confined. For $s<0$, the boson field $z$ is condensed, while the fermionic field $\xi$ cannot be condensed. This spontaneously breaks the $OSp(1|2)$ symmetry since $n_a=\bar{z}\sigma_a z$ is consequently condensed, Higgsing the $U(1)$ gauge field and resulting in the SN phase. From (\ref{eq:30}), the expectation value of $z$ at mean-field level is $\left<|z|\right>=\sqrt{\frac{-s}{2u}}$, which produce a mass counter term $\delta\mathcal{L}=-s\xi\bar{\xi}$ that cancels the symplectic fermion mass $s$. The fermion Lagrangian in the SN phase (with the Higgsed $U(1)$ gauge field omitted),
\beq
\mathcal{L}_{\mathrm{SN}}=\partial_\mu\xi\partial_\mu\bar{\xi}=\partial_\mu\vartheta_\alpha J_{\alpha\beta}\partial_\mu\vartheta_\beta,\label{eq:31}
\eeq
is gapless. In the second equality of (\ref{eq:31}), $\vartheta_{1,2}$ are real and imaginary part of $\xi$, respectively. Thus, $\vartheta_{1,2}$ are exactly the Goldstino modes in the SN phase, as the SUSY partners of Goldstone bosons arising from the fluctuation of SN order parameter $n_a$. By combining the Goldstone and Goldstino modes, we recover the NL$\sigma$M (\ref{eq:14}) in the SN phase with identification $\vartheta_\alpha\sim \eta_\alpha$. Physically, across the critical point, the asymptotically deconfined symplectic fermion in the SVBS phase becomes gapless in the SN phase and plays the role of Goldstino modes.

The internal $OSp(1|2)$ symmetry protects that the boson field $z$ and fermion field $\xi$ must be simultaneously gapless at the critical point. Therefore, the universality class of the sDQCP should be drastically different from the DQCP~\cite{Senthil:2004aa,Senthil:2004ab}. In the literature~\cite{Guruswamy:1998,Read:2001,Jacobsen:2005,Parisi:1980}, critical symplectic fermions are often called {\it negative} degrees of freedom, since they have negative central charges due to their non-unitarity. More precisely, the $-1$ factor in fermion loops of the Feynman diagram cancels the contribution of boson loops~\cite{Parisi:1980}. As a result, 1 complex symplectic fermion degree of freedom can be viewed as $-2$ real boson~\cite{Parisi:1980} or equivalently $-1$ complex boson degrees of freedom. Indeed, in the renormalization group calculations in (1+1)D~\cite{Read:2001,Babichenko2007,Candu2010} and (2+1)D~\cite{LeClair:2007}, symplectic fermions contribute negatively in the $\beta$-function~\cite{Read:2001,Babichenko2007,Candu2010,LeClair:2007}, while bosons contribute positively. In particular, an $N$-flavor symplectic fermion has the same $\beta$-function up to two loop as an $O(M)$ real boson theory with $M=-2N$~\cite{LeClair:2007}, which supports the argument above. By counting of degrees of freedom, at the critical point, the gapless symplectic fermion field $\xi$ cancels one gapless complex boson field, say $z_1$, leaving an effective critical Lagrangian with only $z_2$,
\beq
\mathcal{L}_\mathrm{eff}=\left|\left(\partial_\mu-ia_\mu\right)z_2\right|^2+u\left|z_2\right|^4+\frac{1}{2\kappa}\left(\epsilon_{\mu\nu\rho}\partial_\nu a_\rho\right)^2,\label{eq:32}
\eeq
which is exactly the critical theory of the Abelian Higgs model belonging to the 3D XY universality class. Loop model studies of the $\mathbb{CP}^{1|1}$ model~\cite{Nahum2012} suggest that its critical point is thermodynamically equivalent to 3D XY, namely they share the same singularity in the free energy. Consequently, the critical exponents for the correlation length and the heat capacity of the sDQCP are equal to those of the 3D XY class. However, as the operator content of the $\mathbb{CP}^{1|1}$ model contains an $OSp(1|2)$ multiplet and is different from the XY model, the critical exponents of the order parameter and other physical operators can be different. These critical exponents are in principle measurable numerically in Monte Carlo simulations, e.g., through finite-size scaling of thermodynamic observables, as well as correlation functions of various operators~\cite{Nahum2011}. That means, the sDQCP should be understood as a type of 3D XY* transition point~\cite{Xu:2012}.

\section{Explicitly breaking the internal SUSY}\label{sec:break}

The pseudo-Hermitian Hamiltonian (\ref{eq:9}) can be made Hermitian via operator $P$ defined in (\ref{eq:10}) as $\tilde{H}=PH$. Here $\tilde{H}$ is Hermitian; however, the internal SUSY $OSp(1|2)$ is explicitly broken to spin $SU(2)_S$. As a result, in the NL$\sigma$M (\ref{eq:20}), the fermionic modes $\eta_\alpha$ are gapped out and thus eliminated from the NL$\sigma$M. This reduces the NL$\sigma$M to the $O(5)$ NL$\sigma$M describing the DQCP between the N\'eel phase and the VBS phases~\cite{Senthil:2004aa,Senthil:2004ab,Grover2008}. On the other hand, the critical theory will also be reduced to the $\mathbb{CP}^1$ model describing the DQCP~\cite{Senthil:2004aa}. To see this, consider the critical regime of $\tilde{H}$, where the symplectic fermion is decoupled from the low energy spectrum since it is incompatible with an Hermitian system. Consequently, the low energy degrees of freedom will be the dynamical $U(1)$ gauge field $a_\mu$, and the bosonic spinon field $z$ carrying spin-$\frac{1}{2}$ under $SU(2)_S$. The fermionic part of the theory can be  written as
\beq
\mathcal{L}_F=(\partial_\mu-ia_\mu)\xi(\partial_\mu+ia_\mu)\bar{\xi}+
\left(s+\delta s\right)\xi\bar{\xi}.\label{eq:33}
\eeq
Here $\delta s>0$ makes the symplectic fermion field $\xi$ more massive than the boson $z$, rendering the explicit breaking of the internal SUSY. When the boson is condensed, the mass counter term it generating, $\delta\mathcal{L}=-s\xi\bar{\xi}$, cannot fully cancel the modified fermion mass in (\ref{eq:33}). The symplectic fermion remains gapped across the critical point and in the boson condensed phase, as
\beq
\mathcal{L}^\prime_{\mathrm{SN}}=\partial_\mu\xi\partial_\mu\bar{\xi}+\delta s\xi\bar{\xi}.\label{eq:34}
\eeq
Therefore, the gapped symplectic fermion $\xi$ does not affect the critical property of the critical point, and the universality of such quantum critical point should be the same as the DQCP between N\'eel and VBS phases~\cite{Senthil:2004aa}. In addition, according to (\ref{eq:34}), in the boson condensed phase there are only Goldstone modes and no gapless Goldstino modes, suggesting an ordinary N\'eel phase instead of the SN phase.

\section{Conclusions}\label{sec:con}

In conclusion, we have extended the DQCP paradigm to systems with \emph{internal} supersymmetry, in which the on-site Hilbert space spans a representation of a Lie superalgebra~\cite{Frappat:1996} and the Hamiltonian is invariant under the corresponding Lie supergroup. Focusing on the minimal supersymmetric generalization of spin-$SU(2)$, $OSp(1|2)$, we proposed a supersymmetric deconfined quantum critical point (sDQCP) between a phase that breaks internal $OSp(1|2)$ and a phase that instead breaks lattice rotation symmetry. We developed complementary continuum descriptions: a non-linear sigma model on an appropriate supersphere target space that encodes the symmetry intertwinement, and a gauge theory formulation that captures the dynamical aspects of the transition, including a route to 3D XY criticality. Finally, we showed that explicitly breaking $OSp(1|2)$ down to $SU(2)$ smoothly connects our sDQCP to the conventional DQCP scenario, providing a unified framework for deconfined criticality with and without internal supersymmetry.

Our work opens several directions for future study. 
First, it would be valuable to substantiate the proposed universality class with a more controlled analysis, for instance via an $\epsilon$--expansion, a large-$N$ generalization, or numerical simulations that can directly access the scaling behavior at the sDQCP. 
Second, the gauge-theory description suggests distinctive low-energy signatures associated with supersymmetry, such as correlated bosonic and fermionic critical modes; it would be interesting to identify sharp observables (e.g.\ operator content, anomalous dimensions, and characteristic correlation functions) that can unambiguously distinguish the sDQCP from its non-supersymmetric counterparts. 
Third, while we have focused on the minimal $OSp(1|2)$ case, it is natural to explore generalizations to other internal supergroup symmetries and to classify which symmetry-breaking patterns admit deconfined criticality with symmetry intertwinement. 
Finally, since internal supersymmetry on the lattice is naturally tied to pseudo-Hermitian settings, it is natural to clarify the extent to which the sDQCP can arise in quantum simulator platforms.

\acknowledgments

We acknowledge T. Senthil, Adam Nahum, Biao Lian, Hart Goldman, and Salvatore Pace for helpful discussions. ZQG acknowledges support from Berkeley graduate program. YQW is supported by the JQI postdoctoral fellowship at the University of Maryland.

\bibliography{susy}

\providecommand{\href}[2]{#2}\begingroup\raggedright\begin{thebibliography}{10}

\bibitem{Senthil:2004aa}
T.~Senthil, A.~Vishwanath, L.~Balents, S.~Sachdev and M.P.A.~Fisher,
  \emph{Deconfined quantum critical points},
  \href{https://doi.org/10.1126/science.10918}{\emph{Science} {\bfseries 303}
  (2004) 1490}.

\bibitem{Senthil:2004ab}
T.~Senthil, L.~Balents, S.~Sachdev, A.~Vishwanath and M.P.A.~Fisher,
  \emph{Quantum criticality beyond the landau-ginzburg-wilson paradigm},
  \href{https://doi.org/10.1103/PhysRevB.70.144407}{\emph{Physical Review B}
  {\bfseries 70} (2004) 144407}.

\bibitem{Grover2007}
T.~Grover and T.~Senthil, \emph{Quantum spin nematics, dimerization, and
  deconfined criticality in quasi-1d spin-one magnets},
  \href{https://doi.org/10.1103/PhysRevLett.98.247202}{\emph{Phys. Rev. Lett.}
  {\bfseries 98} (2007) 247202}.

\bibitem{Grover2008}
T.~Grover and T.~Senthil, \emph{Topological spin hall states, charged
  skyrmions, and superconductivity in two dimensions},
  \href{https://doi.org/10.1103/PhysRevLett.100.156804}{\emph{Phys. Rev. Lett.}
  {\bfseries 100} (2008) 156804}.

\bibitem{Wang2015}
F.~Wang, S.A.~Kivelson and D.-H.~Lee, \emph{Nematicity and quantum
  paramagnetism in fese}, \href{https://doi.org/10.1038/nphys3456}{\emph{Nature
  Physics} {\bfseries 11} (2015) 959–963}.

\bibitem{Roberts2019}
B.~Roberts, S.~Jiang and O.I.~Motrunich, \emph{Deconfined quantum critical
  point in one dimension},
  \href{https://doi.org/10.1103/PhysRevB.99.165143}{\emph{Phys. Rev. B}
  {\bfseries 99} (2019) 165143}.

\bibitem{Zhang2023}
C.~Zhang and M.~Levin, \emph{Exactly solvable model for a deconfined quantum
  critical point in 1d},
  \href{https://doi.org/10.1103/PhysRevLett.130.026801}{\emph{Phys. Rev. Lett.}
  {\bfseries 130} (2023) 026801}.

\bibitem{Cui2025}
Y.~Cui, R.~Yu and W.~Yu, \emph{Deconfined quantum critical point: A review of
  progress}, \href{https://doi.org/10.1088/0256-307X/42/4/047503}{\emph{Chin.
  Phys. Lett.} {\bfseries 42} (2025) 047503}.

\bibitem{Xu:2012}
C.~Xu, \emph{{Unconventional} {Quantum} {Critical} {Points}},
  \href{https://doi.org/10.1142/s0217979212300071}{\emph{International Journal
  of Modern Physics B} {\bfseries 26} (2012) 1230007}.

\bibitem{Landau}
L.D.~Landau, E.M.~Lifshits and L.P.~Pitaevskii, \emph{Statistical Physics:
  Theory of the Condensed State, Volume 9}, vol.~9, Pergamon Press (1980).

\bibitem{Wang:2017uq}
C.~Wang, A.~Nahum, M.A.~Metlitski, C.~Xu and T.~Senthil, \emph{Deconfined
  quantum critical points: Symmetries and dualities},
  \href{https://doi.org/10.1103/PhysRevX.7.031051}{\emph{Physical Review X}
  {\bfseries 7} (2017) 031051}.

\bibitem{Metlitski2018}
M.A.~Metlitski and R.~Thorngren, \emph{Intrinsic and emergent anomalies at
  deconfined critical points},
  \href{https://doi.org/10.1103/PhysRevB.98.085140}{\emph{Phys. Rev. B}
  {\bfseries 98} (2018) 085140}.

\bibitem{Lu2023}
D.-C.~Lu, \emph{{Nonlinear sigma model description of deconfined quantum
  criticality in arbitrary dimensions}},
  \href{https://doi.org/10.21468/SciPostPhysCore.6.3.047}{\emph{SciPost Phys.
  Core} {\bfseries 6} (2023) 047}.

\bibitem{Wang2024}
Y.-Q.~Wang, C.~Liu and Y.-M.~Lu, \emph{Theory of topological defects and
  textures in two-dimensional quantum orders with spontaneous symmetry
  breaking}, \href{https://doi.org/10.1103/PhysRevB.109.195165}{\emph{Phys.
  Rev. B} {\bfseries 109} (2024) 195165}.

\bibitem{Pace2024}
S.D.~Pace, \emph{{Emergent generalized symmetries in ordered phases and
  applications to quantum disordering}},
  \href{https://doi.org/10.21468/SciPostPhys.17.3.080}{\emph{SciPost Phys.}
  {\bfseries 17} (2024) 080}.

\bibitem{Sandvik:2007aa}
A.W.~Sandvik, \emph{Evidence for deconfined quantum criticality in a
  two-dimensional heisenberg model with four-spin interactions},
  \href{https://doi.org/10.1103/PhysRevLett.98.227202}{\emph{Physical Review
  Letters} {\bfseries 98} (2007) 227202}.

\bibitem{Kuklov:2008aa}
A.B.~Kuklov, M.~Matsumoto, N.V.~Prokof'ev, B.V.~Svistunov,  and M.~Troyer,
  \emph{Deconfined criticality: Generic first-order transition in the su(2)
  symmetry case},
  \href{https://doi.org/10.1103/PhysRevLett.101.050405}{\emph{Physical Review
  Letters} {\bfseries 101} (2008) 050405}.

\bibitem{li2019}
Z.-X.~Li, S.-K.~Jian and H.~Yao, \emph{Deconfined quantum criticality and
  emergent so(5) symmetry in fermionic systems},  2019.

\bibitem{Huang2019}
R.-Z.~Huang, D.-C.~Lu, Y.-Z.~You, Z.Y.~Meng and T.~Xiang, \emph{Emergent
  symmetry and conserved current at a one-dimensional incarnation of deconfined
  quantum critical point},
  \href{https://doi.org/10.1103/physrevb.100.125137}{\emph{Physical Review B}
  {\bfseries 100} (2019) }.

\bibitem{Zhao2022}
J.~Zhao, Y.-C.~Wang, Z.~Yan, M.~Cheng and Z.Y.~Meng, \emph{Scaling of
  entanglement entropy at deconfined quantum criticality},
  \href{https://doi.org/10.1103/PhysRevLett.128.010601}{\emph{Phys. Rev. Lett.}
  {\bfseries 128} (2022) 010601}.

\bibitem{Liu2023}
Z.H.~Liu, W.~Jiang, B.-B.~Chen, J.~Rong, M.~Cheng, K.~Sun et~al., \emph{Fermion
  disorder operator at gross-neveu and deconfined quantum criticalities},
  \href{https://doi.org/10.1103/physrevlett.130.266501}{\emph{Physical Review
  Letters} {\bfseries 130} (2023) }.

\bibitem{Song2025}
M.~Song, J.~Zhao, M.~Cheng, C.~Xu, M.~Scherer, L.~Janssen et~al.,
  \emph{Evolution of entanglement entropy at su( n ) deconfined quantum
  critical points}, \href{https://doi.org/10.1126/sciadv.adr0634}{\emph{Science
  Advances} {\bfseries 11} (2025) }.

\bibitem{Motrunich:2004}
O.I.~Motrunich and A.~Vishwanath, \emph{Emergent photons and transitions in the
  o(3) sigma model with hedgehog suppression},
  \href{https://doi.org/10.1103/PhysRevB.70.075104}{\emph{Physical Review B}
  {\bfseries 70} (2004) 075104}.

\bibitem{Haldane:1988}
F.D.M.~Haldane, \emph{O(3) nonlinear sigma model and the topological
  distinction between integer- and half-integer-spin antiferromagnets in two
  dimensions},
  \href{https://doi.org/10.1103/PhysRevLett.61.1029}{\emph{Physical Review
  Letters} {\bfseries 61} (1988) 1029}.

\bibitem{Golfand1971}
Y.A.~Golfand and E.P.~Likhtman, \emph{Extension of the algebra of poincare
  group generators and violation of p invariance}, {\emph{Jetp Letters}
  {\bfseries 13} (1971) 323}.

\bibitem{Wess1974}
J.~Wess and B.~Zumino, \emph{Supergauge transformations in four dimensions},
  \href{https://doi.org/https://doi.org/10.1016/0550-3213(74)90355-1}{\emph{Nuclear
  Physics B} {\bfseries 70} (1974) 39}.

\bibitem{Freedman1976}
D.Z.~Freedman, P.~van Nieuwenhuizen and S.~Ferrara, \emph{Progress toward a
  theory of supergravity},
  \href{https://doi.org/10.1103/PhysRevD.13.3214}{\emph{Phys. Rev. D}
  {\bfseries 13} (1976) 3214}.

\bibitem{Deser1976}
S.~Deser and B.~Zumino, \emph{Consistent supergravity},
  \href{https://doi.org/https://doi.org/10.1016/0370-2693(76)90089-7}{\emph{Physics
  Letters B} {\bfseries 62} (1976) 335}.

\bibitem{Fayet1976}
P.~Fayet, \emph{Supersymmetry and weak, electromagnetic and strong
  interactions},
  \href{https://doi.org/https://doi.org/10.1016/0370-2693(76)90319-1}{\emph{Physics
  Letters B} {\bfseries 64} (1976) 159}.

\bibitem{Fayet1977}
P.~Fayet, \emph{Spontaneously broken supersymmetric theories of weak,
  electromagnetic and strong interactions},
  \href{https://doi.org/https://doi.org/10.1016/0370-2693(77)90852-8}{\emph{Physics
  Letters B} {\bfseries 69} (1977) 489}.

\bibitem{Sakai1981}
N.~Sakai, \emph{Naturalnes in supersymmetric guts},
  \href{https://doi.org/10.1007/BF01573998}{\emph{Zeitschrift f{\"u}r Physik C
  Particles and Fields} {\bfseries 11} (1981) 153}.

\bibitem{Fendley:2003aa}
P.~Fendley, K.~Schoutens and
  J.~de~Boer\href{https://doi.org/10.1103/PhysRevLett.90.120402}{\emph{Physical
  Review Letters} {\bfseries 90} (2003) 120402}.

\bibitem{Yang:2004}
X.~Yang and P.~Fendley, \emph{Non-local spacetime supersymmetry on the
  lattice}, \href{https://doi.org/10.1088/0305-4470/37/38/003}{\emph{Journal of
  Physics A: Mathematical and General} {\bfseries 37} (2004) 8937}.

\bibitem{OBrien:2018aa}
E.~O'Brien and P.~Fendley, \emph{Lattice supersymmetry and order-disorder
  coexistence in the tricritical ising model},
  \href{https://doi.org/10.1103/PhysRevLett.120.206403}{\emph{Physical Review
  Letters} {\bfseries 120} (2018) }.

\bibitem{Gao:2020}
Z.-Q.~Gao and C.~Wu, \emph{Construction of $g_2$ symmetry in a hubbard-type
  model},  2024.

\bibitem{Gao2024}
Z.-Q.~Gao and C.~Wu, \emph{From g2 to so(8): Emergence and reminiscence of
  supersymmetry and triality},
  \href{https://doi.org/10.1007/JHEP02(2025)202}{\emph{Journal of High Energy
  Physics} {\bfseries 2025} (2025) 202}.

\bibitem{Frappat:1996}
L.~Frappat, P.~Sorba and A.~Sciarrino, \emph{{Dictionary on Lie
  superalgebras}},  \href{https://arxiv.org/abs/hep-th/9607161}{{\ttfamily
  hep-th/9607161}}.

\bibitem{Arovas:2009}
D.P.~Arovas, K.~Hasebe, X.-L.~Qi and S.-C.~Zhang, \emph{Supersymmetric valence
  bond solid states},
  \href{https://doi.org/10.1103/PhysRevB.79.224404}{\emph{Physical Review B}
  {\bfseries 79} (2009) 224404}.

\bibitem{Hasebe:2013}
K.~Hasebe and K.~Totsuka, \emph{Topological many-body states in quantum
  antiferromagnets via fuzzy supergeometry},
  \href{https://doi.org/10.3390/sym5020119}{\emph{Symmetry} {\bfseries 5}
  (2013) 119}.

\bibitem{Saleur:2003}
H.~Saleur and B.~Wehefritz-Kaufmann, \emph{Integrable quantum field theories
  with supergroup symmetries: the osp(1/2) case},
  \href{https://doi.org/https://doi.org/10.1016/S0550-3213(03)00385-7}{\emph{Nuclear
  Physics B} {\bfseries 663} (2003) 443}.

\bibitem{Frahm:2022}
H.~Frahm and M.J.~Martins, \emph{Osp(n|2m) quantum chains with free
  boundaries},
  \href{https://doi.org/https://doi.org/10.1016/j.nuclphysb.2022.115799}{\emph{Nuclear
  Physics B} {\bfseries 980} (2022) 115799}.

\bibitem{Pauli:1943}
W.~Pauli, \emph{On dirac's new method of field quantization},
  \href{https://doi.org/10.1103/RevModPhys.15.175}{\emph{Reviews of Modern
  Physics} {\bfseries 15} (1943) 175}.

\bibitem{Bender:1999}
C.M.~Bender, S.~Boettcher and P.N.~Meisinger, \emph{{PT}-symmetric quantum
  mechanics}, \href{https://doi.org/10.1063/1.532860}{\emph{Journal of
  Mathematical Physics} {\bfseries 40} (1999) 2201}.

\bibitem{Mostafazadeh:2002}
A.~Mostafazadeh, \emph{Pseudo-hermiticity versus {PT} symmetry: The necessary
  condition for the reality of the spectrum of a non-hermitian hamiltonian},
  \href{https://doi.org/10.1063/1.1418246}{\emph{Journal of Mathematical
  Physics} {\bfseries 43} (2002) 205}.

\bibitem{Rabin:1985}
J.M.~Rabin and L.~Crane, \emph{Global properties of supermanifolds},
  \href{https://doi.org/10.1007/bf01212690}{\emph{Communications in
  Mathematical Physics} {\bfseries 100} (1985) 141}.

\bibitem{Hasebe:2005}
K.~Hasebe and Y.~Kimura, \emph{Fuzzy supersphere and supermonopole},
  \href{https://doi.org/https://doi.org/10.1016/j.nuclphysb.2004.11.040}{\emph{Nuclear
  Physics B} {\bfseries 709} (2005) 94}.

\bibitem{CFT}
P.D.~Francesco, P.~Mathieu and D.~S{\'{e}}n{\'{e}}chal, \emph{Conformal Field
  Theory}, Springer New York (1997),
  \href{https://doi.org/10.1007/978-1-4612-2256-9}{10.1007/978-1-4612-2256-9}.

\bibitem{Huang:2023}
Y.-T.~Huang and D.-H.~Lee, \emph{Competing orders, the wess-zumino-witten term,
  and spin liquids},
  \href{https://doi.org/https://doi.org/10.1016/j.nuclphysb.2022.116043}{\emph{Nuclear
  Physics B} {\bfseries 986} (2023) 116043}.

\bibitem{Read:2001}
N.~Read and H.~Saleur, \emph{Exact spectra of conformal supersymmetric
  nonlinear sigma models in two dimensions},
  \href{https://doi.org/https://doi.org/10.1016/S0550-3213(01)00395-9}{\emph{Nuclear
  Physics B} {\bfseries 613} (2001) 409}.

\bibitem{Kurkcuoglu:2004}
S.~Kurkcuoglu, \emph{Non-linear sigma model on the fuzzy supersphere},
  \href{https://doi.org/10.1088/1126-6708/2004/03/062}{\emph{Journal of High
  Energy Physics} {\bfseries 2004} (2004) 062}.

\bibitem{Landi:2001}
G.~Landi, \emph{Projective modules of finite type over the supersphere s2,2},
  \href{https://doi.org/https://doi.org/10.1016/S0926-2245(00)00041-3}{\emph{Differential
  Geometry and its Applications} {\bfseries 14} (2001) 95}.

\bibitem{LeClair:2006}
A.~LeClair, \emph{Quantum critical spin liquids and conformal field theory in
  2+1 dimensions},  2007.

\bibitem{LeClair:2007}
A.~LeClair and M.~Neubert, \emph{Semi-lorentz invariance, unitarity, and
  critical exponents of symplectic fermion models},
  \href{https://doi.org/10.1088/1126-6708/2007/10/027}{\emph{Journal of High
  Energy Physics} {\bfseries 2007} (2007) 027}.

\bibitem{Kapit:2009}
E.~Kapit and A.~LeClair, \emph{Non-fermi liquid properties of 2d symplectic
  fermions: the role of a dynamically generated (pseudo)-gap}, {\emph{arXiv}
  (2009) 0903.2484}.

\bibitem{Robinson:2009}
D.J.~Robinson, E.~Kapit and A.~LeClair, \emph{Lorentz symmetric quantum field
  theory for symplectic fermions},
  \href{https://doi.org/10.1063/1.3248256}{\emph{Journal of Mathematical
  Physics} {\bfseries 50} (2009) 112301}.

\bibitem{Guruswamy:1998}
S.~Guruswamy and A.W.~Ludwig, \emph{Relating $c<0$ and $c>0$ conformal field
  theories},
  \href{https://doi.org/https://doi.org/10.1016/S0550-3213(98)00059-5}{\emph{Nuclear
  Physics B} {\bfseries 519} (1998) 661}.

\bibitem{Jacobsen:2005}
J.L.~Jacobsen and H.~Saleur, \emph{The arboreal gas and the supersphere sigma
  model},
  \href{https://doi.org/https://doi.org/10.1016/j.nuclphysb.2005.04.001}{\emph{Nuclear
  Physics B} {\bfseries 716} (2005) 439}.

\bibitem{Parisi:1980}
G.~Parisi and N.~Sourlas, \emph{Self avoiding walk and supersymmetry},
  \href{https://doi.org/10.1051/jphyslet:019800041017040300}{\emph{Journal de
  Physique Lettres} {\bfseries 41} (1980) 403}.

\bibitem{Babichenko2007}
A.~Babichenko, \emph{Conformal invariance and quantum integrability of sigma
  models on symmetric superspaces},
  \href{https://doi.org/10.1016/j.physletb.2007.03.003}{\emph{Physics Letters
  B} {\bfseries 648} (2007) 254–261}.

\bibitem{Candu2010}
C.~Candu, V.~Mitev, T.~Quella, H.~Saleur and V.~Schomerus, \emph{The sigma
  model on complex projective superspaces},
  \href{https://doi.org/10.1007/jhep02(2010)015}{\emph{Journal of High Energy
  Physics} {\bfseries 2010} (2010) 015}.

\bibitem{Nahum2012}
A.~Nahum and J.T.~Chalker, \emph{Universal statistics of vortex lines},
  \href{https://doi.org/10.1103/PhysRevE.85.031141}{\emph{Phys. Rev. E}
  {\bfseries 85} (2012) 031141}.

\bibitem{Nahum2011}
A.~Nahum, J.T.~Chalker, P.~Serna, M.~Ortu\~no and A.M.~Somoza, \emph{3d loop
  models and the ${\mathrm{cp}}^{n\ensuremath{-}1}$ sigma model},
  \href{https://doi.org/10.1103/PhysRevLett.107.110601}{\emph{Phys. Rev. Lett.}
  {\bfseries 107} (2011) 110601}.

\end{thebibliography}\endgroup
\bibliographystyle{JHEP}

\end{document}